\documentclass[12pt]{article}
\newtheorem{prop}{Proposition}[section]
\newtheorem{property}{Property}
\newtheorem{remark}{Remark}
\input{psfig}   
\usepackage{epsfig} 
\usepackage{latexsym}
\title{Contact between laboratory instruments and \\ equations of
quantum mechanics} 
\author{John M. Myers\\Gordon McKay Laboratory, Harvard University,\\
Cambridge, MA 02138, USA \\ \\  and  \\ \\ F. Hadi Madjid \\
82 Powers Road, Concord, MA 01742, USA }

\begin{document}  \maketitle  
\begin{abstract}

Ambiguity in the contact between laboratory instruments and equations
of quantum mechanics is formulated in terms of responses of the
instruments to commands transmitted to them by a Classical digital
Process-control Computer (CPC); in this way instruments are
distinguished from quantum-mechanical models (sets of equations) that
specify what is desired of the instruments.  Results include:
\begin{enumerate} \item a formulation of quantum mechanics adapted to
computer-control\-led instruments; \item a lower bound on the precision
of unitary transforms required for quantum searching and a lower bound
on sample size needed to show that instruments implement a desired
model at that precision; \item a lower bound on precision of timing
required of a CPC in directing instruments; \item a demonstration that
guesswork is necessary in ratcheting up the precision of
commands. \end{enumerate}
\end{abstract}


\section{Introduction}
\label{sect:intro}  
 
To build a quantum computer is to arrange for laboratory instruments
to produce results in accord with models expressed in equations of
quantum mechanics.  These quantum-mechanical models are of two types:
\begin{enumerate} \item instrument-independent models focused on
relating the multiplication of unitary operators to the solving of
problems of interest, and \item models that tie the multiplication of
unitary operators (as well as steps of state preparation and
measurement) to the use of laboratory instruments. \end{enumerate}
Instruments that could implement a quantum computer would be valuable
if their results could substitute for a more costly classical
calculation defined by a model of type 1.  If a scientist has such
instruments, however, using them requires a model of type 2
which tells the scientist how to use the instruments; it is these
quantum-mechanical models of how instruments function that are the
subject of this report.

Models of type 2 are subject to revision in the course of using
instruments, as we found in arranging for a nuclear-magnetic-resonance
(NMR) spectrometer to implement a model of an NMR quantum computer.
In addition to its electromagnet that holds a liquid sample, the
spectrometer has a Classical, digital Process-control Computer (CPC),
a variable radio-frequency (r.f.)  transmitter controlled by the CPC,
and an r.f. receiver that reports back to the CPC which then computes
spectra and displays them.  Our colleagues synthesized the active
constituent of the liquid for a 5-spin quantum computer, aiming to
have the spectrometer behave roughly as described by a certain model
$\alpha$ (of type 2), characterized by a Hamiltonian for 5
linearly-coupled spin-active nuclei in the presence of the variable
r.f. field.\cite{marx2} But while monitoring the chemical synthesis
they found spectra that better fit a model $\beta$ (also of type 2)
characterized by a more complex hamiltonian with one spin-spin
coupling weaker than desired and with extra couplings beyond those
wanted.  So model $\beta$ rather than the original model $\alpha$ was
used to design commands by which to operate the 5-spin quantum
computer, showing that implementers have to negotiate with the
instruments to refine their models.

Another lessen from the 5-spin endeavor is the secondary role of
fields and particles in the design of instruments.  We need models of
how the spectrometer works to tell us what commands the CPC must issue
to the transmitter in order to solve the Deutsch-Jozsa problem, and
the historically available constructs and examples from which to
construct these models are fields, particles and their couplings;
these however are needed only as pieces from which to compose models
to describe the instruments.  A change from model $\alpha$ to model
$\beta$ changes the couplings, and more severe changes change the
particles and fields.

Is the need for alternative models serious?  Whether one is working
with a mathematical model or working with instruments, it is easy to
assume that what goes on in the model, with its fields and particles,
also goes on in the instruments, at least once the ``right'' model is
found.  But recently it was proved that multiple, inequivalent models
to describe a set-up of instruments contend for acceptance not just in
the early stages of an experiment but throughout.\cite{ams} By
assuring many models and hence many configurations of particles and
fields to describe a single set-up of instruments, this proof
punctures the idea that particles and fields (or, more generally,
states and operators) reside in instruments, somehow uniquely situated
in them, if only one could see them.  This puncture relieves a
widespread confusion between models and instruments by demonstrating a
certain independence between the two.  Given that independence, one
has the question: of what does contact between instruments and models
consist?

For purposes of an analysis that recognizes their independence, we
focus on the contact between quantum-mechanical equations and
laboratory instruments that takes place in computer files of a
Classical, digital Process-control Computer (CPC).  Within a session
in which both equation writing and the use of instruments are mediated
by the CPC, a scientist uses the CPC to: \begin{itemize} \item compute
(classically!) with various sets of equations of quantum mechanics
that define what the instruments are supposed to do, and \item send
commands to instruments intended to implement those equations and
record results produced by the instruments. \end{itemize}

Quantum mechanical models appropriate to the CPC control of
instruments are formulated Section 2.  Section 3 addresses the
a question of finding models from which to determine commands
for any particular set-up of instruments, leading to an analysis of
sample sizes of experimental tests of models as a function of
precision.  Consistent with prior estimates\cite{vaz}, in
quantum searching\cite{grover} the required sample size rises
exponentially with the number of bits.

Section 4 shows lower bounds on the precision of timing needed for a
CPC to manage quantum-computing instruments.  Section 5 introduces
the concept of a lattice of models to show that testing in physics and
engineering cannot be universal like the test of a derivation in
mathematics; instead, a scientist working with instruments, trying for
progressively higher precision in their accord with models, must
re-evaluate properties by which to restrict a set of models, requiring
repeated resort to guesswork.

\section{Quantum mechanical models for CPC-controlled instruments}

For one example of contact between instruments and equations, suppose a
scientist has instruments which would implement a quantum computer if
they were sent correct commands; the scientist then faces the question
of what commands the CPC should transmit and when it should transmit
to them in order {\em e.g.} to implement a quantum gate expressed as a
unitary matrix $U_j$.

To produce a core set of models from which a scientist can choose to
describe the working of particular instruments, we suppose that during
a CPC-mediated session some instruments are controlled by
CPC-transmitted commands from a set $B$ of possible commands, where $B
\subset \mathcal{B}$ and $\mathcal{B}$ is the set of all finite binary
strings.  Our models express the probability of an outcome of
instruments in response to a command $b \in B$ sent to the instruments
by the CPC, as follows.  Let $\mathcal{H}$ be a separable Hilbert
space.  Let $\mathcal{V}_B$, $\mathcal{U}_B$, and $\mathcal{M}_B$ be
the sets of all functions $|v\rangle$, $U$, and $M$, respectively,
with \begin{eqnarray*} |v\rangle: B & \rightarrow & \mathcal{H},\\ U:
B & \rightarrow & \{\mbox{unitary operators on } \mathcal{H} \},\
\mbox{and}\\ M: B & \rightarrow & \{\mbox{hermitian operators on }
\mathcal{H} \}. \end{eqnarray*} The core models exhibit discrete
spectra for all $M \in \mathcal{M}$: \begin{property}
\begin{equation} (\forall b \in B) M(b) = \sum_j m_j(b) M_j(b),
\label{eq:specthm} \end{equation} where $m_j: B \rightarrow \mathcal{R}$
(with $\mathcal{R}$ denoting the real numbers) is the $j$-th
eigenvalue of $M$, and $M_j$ is the projection onto the $j$-th
eigenspace (so $M_j M_k = \delta_{j,k}M_j$).
\end{property} Let $\Pr(j|b)$ denote the probability of obtaining the
$j$-th outcome, given transmission by the CPC of a command $b$.
Although not commonly seen in texts, this probability of an outcome
given a command is the hinge pin for focusing on quantum mechanical
language used to describe what a scientist can find by using
instruments.  Quantum mechanics constrains all these models to satisfy
\begin{property} \begin{equation} \Pr(j|b) = \langle
v(b)|U^\dagger(b)M_j(b)U(b)|v(b)\rangle , \end{equation} where the
$\dagger$ denotes the hermitian adjoint. \label{eq:basic}
\end{property}

Any choice of command set $B$ and of functions from the sets
$\mathcal{V}_B$, $\mathcal{U}_B$, and $\mathcal{M}_B$ produces a
quantum-mechan\-ical model $(|v\rangle, U, M)_B$.  Two models
$(|v\rangle, U, M)_B$ and $(|v'\rangle, U', M')_B$ generate the same
probabilities $\Pr(j|b)$ if they are unitarily equivalent, meaning
there exists a $Q: B \rightarrow \{\mbox{unitary operators on }
\mathcal{H} \}$ such that $(\forall b \in B) |v'(b)\rangle =
Q(b)|v(b)\rangle$, $U'(b) = Q(b) U(b) Q^\dagger(b)$ and $M'(b) = Q(b)
M(b) Q^\dagger(b)$.  For this reason, any model $(|v\rangle, U, M)_B$
can be reduced to $(|v'\rangle, {\bf 1}, M)_B$, where $|v'\rangle =
U|v\rangle$ and $M' = M$ or, alternatively to $(|v\rangle, {\bf 1},
M')_B$ where $M' = U^\dagger M U$.

As was recently proved\cite{ams}, in order that measured data can
select a single best fitting model from a set of models (up to unitary
equivalence), additional restrictions are necessary to narrow down the
set of models to a much smaller set than that defined by properties 1
and 2, because many inequivalent models can be found to fit exactly
and conceivable record of outcomes.  Furthermore, it was proved that
these restrictions cannot be derived from quantum mechanics nor from
the measured data, so that imposing them takes guesswork on the part
of the scientist.  This core set of models is a subset of models
available using more general formulations.  Because guesswork is
necessary even to resolve choices among models of the core
set\cite{ams}, it follows that guesswork is necessary also to resolve
the broader choices of models from supersets that include more models,
{\em e.g.} models involving positive-operator-valued measures or
superoperators or both.  

It is conventional in endeavors aiming at quantum computing to assume
three additional properties to narrow the set of models; this is
community-endorsed guesswork:

\begin{property} The command $b$ is the
concatenation of separate commands for the three types of operations, so
that
\begin{equation}
b = b_v\parallel b_U\parallel b_M \label{eq:bcat}, \end{equation}
where here the $\parallel$ denotes concatenation of commands.
\end{property}
According to these models, one can vary any one of the three while
holding the other two fixed.  This specializes Eq.\ (\ref{eq:basic})
to the more restrictive form:
\begin{equation}
\Pr(j|b) = \langle v(b_v)|U^\dagger(b_U)M_j(b_M)U(b_U)|v(b_v)\rangle .
\label{eq:basic2} \end{equation}

An additional constraining guess characterizes models widely used in
the analysis of quantum computers, a guess prompted by the desire to
generate a unitary transformation as a product of other unitary
transformations that serve as ``elementary quantum gates.''  For
example, a scientist may want to generate the unitary transformation
$U(b_{U,1})U(b_{U,2})$ by causing the CPC to transmit
some $b_U$.  For quantum computing to have an advantage over classical
computing, the determination of this $b_U$ in terms of $b_{U,1}$ and
$b_{U,2}$ must be of polynomial complexity \cite{complex}.  It is
usually assumed that $b_U$ is the simplest possible function of
$b_{U,1}$ and $b_{U,2}$, as follows.

Let $B_U \subset B$ be a set of instrument-controlling commands,
thought of as strings that can be concatenated.  Suppose the function
$U$ has the form $U(b_1\parallel b_2) = U(b_2)U(b_1)$ for all $b_1
\parallel b_2 \in {B_U}$ (note reversal of order).  Then we say the
function $U$ respects concatenation.

\begin{property} Quantum computation employs a subset of models in
which $U$ respects concatenation.  \end{property}

Finally, the theories widely used in quantum computing assume 
something about timing:
\begin{property} the
unitary transformation commanded by any command $b_U$s takes a state
$|v\rangle$ at one time into a a state $|v' \rangle$ at a time later
by some amount $T(b_U)$.
\end{property}

\begin{remark} To appreciate contact between instruments and equations,
it is essential to see Properties 1 through 5 not as properties of
laboratory instruments, but as properties that a scientist can choose
to demand of models.  Whether the instruments act that way is another
question.  There are reasons, relaxation and other forms of
decoherence among them, to expect limits to the precision with which
instruments can behave in accord with properties 3 through 5.  All
five properties are used often enough to be conventions, in the sense
that a convention is a guess endorsed by a community.  \end{remark}

\subsection{Statistically significant differences between models}

In practice, a scientist has little interest in a model chosen so that
its probabilities exactly fit measured relative frequencies.  Rather,
the scientist wants a simpler model with some appealing structure that
comes reasonably close to fitting.  Quantum mechanics encourages this
predilection, because on account of statistical variation in the
sample mean, functions that perfectly fit outcomes on hand at one time
are not apt to fit perfectly outcomes acquired subsequently.  We
show here that accepting statistics no way takes away from the proof
that measurements and equations by themselves cannot link models to
instruments.

One needs a criterion for the statistical significance of a difference
between two quantum-mechanical models (or between a model and measured
relative frequencies).  Here we limit our attention to models $\alpha$
and $\beta$ which have a set $B$ of commands in common and for which
the spectra of $M_\alpha$ and $M_\beta$ are the same.  For a single
command $b$, the question is whether the difference between the
probability distributions $\Pr_\alpha(\cdot|b)$ and
$\Pr_\beta(\cdot|b)$ is bigger than typical fluctuations expected in
$N(b)$ trials.  An answer is that two distributions are
indistinguishable statistically in $N(b)$ trials unless
\begin{equation} N(b)^{1/2}d(\mbox{$\Pr_\alpha$}(\cdot|b),
\mbox{$\Pr_\beta$}(\cdot|b)) > 1, \label{eq:dist} \end{equation} where
$d$ is the statistical distance defined in Eq.\ (10) of a paper by
Wooters\cite{wooters}.  Furthermore, Wooters's Eq.\ (12) shows for two
models $\alpha$ and $\beta$ that differ only in the function
$|v\rangle$,
\begin{equation} d(\mbox{$\Pr_\alpha$}(\cdot|b),
\mbox{$\Pr_\beta$}(\cdot|b)) \leq \cos^{-1}|\langle v_\alpha(b)|
v_\beta(b) \rangle |. \label{eq:dangle}
\end{equation}

To help judge the significance of the difference between two models
with respect to a set B of commands common to them, a scientist who
chooses some weighting of different commands can define a weighted
average of $d(\Pr_\alpha(\cdot|b), \Pr_\beta(\cdot|b))$ over all $b
\in B$.  The same holds if model $\beta$ is replaced by relative
frequencies of outcomes interpreted from measured results.

It has been proved that the set of models statistically
indistinguishable from a given model can be much larger than
would be the case if the ``$\leq$'' of (\ref{eq:dangle}) were an
equality\cite{ams}.

\begin{prop} For any set of outcomes, two models $\alpha$ and $\beta$ 
of the form $(|v\rangle,{\bf 1},M)_B$ can perfectly fit the relative
frequencies of the outcomes (Proposition 2.1) and yet be mutually
orthogonal in the sense that $\langle v_\alpha| v_\beta \rangle = 0$
\end{prop}

Wooters extended the definition of statistical difference to unit
vectors representing quantum states.  While for any two unit vectors,
there exist measurement operators that maximize the statistical
distance between them, for any such operator there exist other
vectors, mutually orthogonal, that have zero statistical distance
relative to this operator.  For this reason, among others, statistics
still leaves the scientist needing something beyond calculation and
measurement to determine a model, for the set of models closer than
$\epsilon$ in weighted statistical distance to certain measured
results certainly includes all the models that exactly fit the data
and, without special restrictions dependent on guesses, this set
includes models that are mutually orthogonal.  Models close to given
measured data are not necessarily close to each other in the
predictions they make.

\section{Large sample size needed to decide which model better fits measured data}

By recognizing that models are distinct from instruments, it is easy
to ask a question of the form: suppose model $\alpha$ assumed to
describe instruments, is wrong and another model, call it $\beta$,
describes the instruments better with respect to outcomes in some
situation.  Even if instruments behave in accord with a model
$\alpha$, it can be a lot of work to show this, and without showing
it, one does not know it.  While there are many models having
properties 1 and 2 that fit any given outcomes of instruments exactly,
making the the models indistinguishable with respect to fit, there are
many other models that do not fit.  Given a model $\alpha$ that fits
and a model $\gamma$ that does not, it was shown\cite{ams} that models
$\alpha$ and $\gamma$ are statistically distinguishable with respect
to their fit to data measured for a command $b$ only if the
statistical distance between $\alpha$ and $\gamma$, which we call
$\epsilon$, is big enough in relation to sample size $N(b)$;
distinguishability requires
\begin{equation} N(b) \geq \epsilon^{-2}. \end{equation}
This raises the question of the size of the statistical distance
$\epsilon$.  Very small values of $\epsilon$ are required to implement
quantum searching\cite{grover}, implying large sample size, as
follows.  The task of searching is often expressed as the task of
finding the value (assumed to exist and to be unique) for which a
binary function $f$ of $n$ bits is 1.  The function $f$ is expressed
by an oracle, represented by a unitary transformation $U_f$, diagonal
in the computational basis of dimension $N = 2^n$, with 1 for each
value except for $-$1 in the place for the argument for which $f$
takes the value 1\cite{collins}.  For the example of the function
$f_0$ such that $f(0) = 1$ and $f(x) = 0$ for $x \neq 0$, $U_f =
U_0 = {\bf 1} - 2|0\rangle \langle 0 |$, The algorithm for quantum
search can be viewed as implementing a model that calls for three
steps, (the second of which is repeated many times):
\begin{enumerate}
\item prepare a state \begin{equation} |w \rangle \stackrel{\rm
def}{=} N^{-1/2}\sum_{j=0}^{N-1} |j\rangle
\end{equation} \item on the order $N^{1/2}$ times, apply $U_f$
followed by $U_w = {\bf 1} - 2 |w \rangle \langle w|$; and \item make
a nondegenerate measurement diagonal in the computational
basis. \end{enumerate} This works because $U_w$ reflects about
the hyperplane perpendicular to $|w \rangle$, $U_f$ reflects
about the hyperplane perpendicular to the basis vector corresponding
to the special value (which is $|0 \rangle$ for the case $f = f_0$),
and the product of the two reflections acts as a rotation moving the
starting vector $|w \rangle$ toward the special vector for $f$.

Now we show explicitly how a small error in a unitary transformation
can cause a failure.  Suppose that the instruments behave not in
accord with the model $\alpha$ described but in accord with a model
$\beta$ that differs from $\alpha$ in that in place of $U_w$, the
transformation is $U_{\tilde{w}}$ which lacks the component $|0
\rangle$:
\begin{equation} |\tilde{w} \rangle \stackrel{\rm def}{=}
(N-1)^{-1/2}\sum_{j=1}^{N-1} |j\rangle . \end{equation} It is easy to
check that the angle $\theta$ between $|w \rangle$ and $| \tilde{w}
\rangle$ defined by $\theta = \cos^{-1}|\langle w | \tilde{w} \rangle
|$ is given by \begin{equation} \theta = cos^{-1}[(1-1/N)^{1/2}]
\approx N^{-1/2} = 2^{-n/2} . \end{equation} Correspondingly, the
error between the desired unitary transformation $U_w$ and the
transformation $U_{\tilde{w}}$ that describes what the instruments do
is readily calculated to be \begin{equation} \epsilon = \parallel U_w
- U_{\tilde{w}} \parallel = 2|\sin \theta | \approx 2 N^{-1/2} =
2^{1-n/2},
\end{equation} which is exponentially small in the number $n$ of
bits.  Though exponentially small, the error $\epsilon$ completely
destroys the quantum computation for the case of $f_0$, because (by
construction) $\langle 0 | \tilde{w} \rangle = 0$, with the result
that
\begin{equation} (U_{\tilde{w}} U_0)^2 = {\bf 1} , \end{equation}
so the repeated applications of $U_{\tilde{w}}U_0$ (in step 2, above)
accomplish nothing, and no magnification of the desired component $|0
\rangle$ is achieved, so with an error in $U_w$ as in Eq. (11),
the outcome has negligible probability of corresponding to the right
answer.

\subsection{Lower bound on sample size required to verify performance}
Exploring contact between instruments and equations requires not just
the recognition of multiple models for given instruments, but also
modeling the commands sent to instruments.  To verify that a command
$b_U$ generates statistics corresponding to a desired unitary matrix
$U_1$ requires trying the command out in conjunction with related
commands $b_{v,j}$ to prepare state vectors ($j$ ranges over a
sufficiently large set).  The number of these vectors must be
larger than the dimension of the vector subspace relevant to problems
in which $U_1$ is used.  This means for $n$-bit quantum computing the
number of vectors is greater than $N = 2^n$.  For each vector, it
follows from Eqs. (7) and (11) that the sample size required to show
that a command $b_U$ produces statistics consistent with any model to
within the precision required for quantum searching is large indeed,
namely $N(b_U) > \epsilon^{-2} > 2^{n-2}$, whence it follows that the
number of trials (the sample size) needed to verify experimentally that
command $b_U$ accords within $\epsilon$ with a model that says it
generates $U_1$ is greater than $2^n N(b_U)$, which in turn is greater
than $2^{2n-2}$.  With less than this amount of testing, the
likelihood that quantum computing instruments will perform to the
required precision is essentially zero.

\section{High precision required for timing of CPC commands}

While within some margin, the physics of a classical computer is
insulated from its manipulating of symbols, that physics is not
insulated from instruments commanded by a CPC; indeed the precision of
timing of the CPC matters critically to the successful functioning of
instruments that it commands.  Per property 5 of section 2, each
unitary transformation takes a state at an earlier time to a state at
a later time.  Thus a unitary transformation happens not all at once,
but over a time duration that depends on how the instruments implement
the transformation.  A written command $b_U$ acts as a musical score.
Like sight reading at a piano, executing a program containing the
command $b_U$ requires converting the character string $b_U$---the
score---into precisely timed actions---like the striking of keys at
the piano.  In this analogy, the piano keys, so to speak, include the
output buffers that control the amplitude, phase, frequency, and
polarization of lasers of an ion-trap quantum computer or of
radio-frequency transmitters for a nuclear-magnetic-resonance (NMR)
quantum computer.

For this reason, executing a command $b_U$ requires parsing it into
pieces (signals) and sending each signal at its proper time, the
specification of which is contained in the string $b_U$.  Either the
CPC that executes a program in which $b_U$ is written parses the
command into signals and transmits each signal at its appointed time,
or the instruments receiving the command $b_U$, unparsed, contain
programmable counters operating in conjunction with a clock that do
this timed parsing.  Such programmable counters themselves constitute
a special-purpose CPC.  So either the scientist's CPC must execute
commands by issuing an appropriately timed sequence of signals, or
some other CPC attached to the instruments must do this.  Either way,
the capacity to execute programmed motion in step with a clock is a
requirement for a CPC, distinct from and in addition to requirements
to act as a Turing machine.

Implicit in models used in quantum computing is an additional
property, to do with the effect of the mistiming of signals
transmitted by a CPC executing a command $b_U$ for a unitary
transformation.  The effect of mistiming is to generate some
(unwanted) command $\tilde{b}_U$ in place of $b_U$.  To think about
the effect of mistiming, one can start with the simple case of a
mistimed NOT-gate for one bit.  Two such gate operations concatenated
result in the identity, which in the Bloch picture is a rotation by $2
\pi$.  If it takes $T$(NOT) seconds to perform the NOT-gate, there is
an angular rotation rate (in radians/s) of $\omega =
\pi/T(\mbox{NOT})$. To avoid making an error greater than $\epsilon$
in the angle of the state vector, it is then necessary to have the
error $\Delta T$ allowable for the CPC in the timing of signals
satisfy:
\begin{equation} \Delta T / T(\mbox{NOT}) \leq \epsilon/\omega 
T(\mbox{NOT}) = \epsilon / \pi . \end{equation} 
For the search algorithm, Eq. (11) implies
\begin{equation} \Delta T /T(\mbox{NOT}) < 2^{-n/2}. \end{equation}
The best precision of timing available from hydrogen masers is
something like 1 part in 10$^{15}$, which fails to be adequate if $ n
> 30 \ln 10/ \ln 2 < 100$; i.e.  the precision of the best hydrogen
masers is less than the precision of timing needed if $n > 99$.
 
\section{Modeling calls for guesswork}
In this section we review propositions proved elsewhere concerning
guesswork required to link quantum-mechanical models to instruments
and contribute a new one; in addition, we describe how the set of
models forms a lattice that a scientist navigates in searching for a
model that fits data to within some prescribed weighted statistical
distance; finally we describe guesswork needed to replace a model that
fits the behavior of instruments at one level of precision with a
model that fits more precisely.

\subsection{Trying for a model that fits the instruments}

Consider a scientist searching for a model $\alpha$ within a weighted
statistical distance $\epsilon$ of relative frequencies of outcomes
obtained by use of the instruments.  The scientist needs to choose a
model from some large set $S$ of models.  To illustrate what is
involved, suppose $S$ is the set of models of section 2, constrained
only to exhibit properties 1 and 2.  As proved earlier\cite{ams},
we have the following three propositions:
\begin{prop} Given any recorded counts of measured outcomes associated
with any set $B$ of commands, the set of models satisfying properties
1 and 2 contains many unitarily inequivalent models $(|v\rangle, U,
M)_B$, each of which fits perfectly the relative frequencies of
outcomes. \end{prop}
\begin{prop} For any set of outcomes, two models $\alpha$ and $\beta$ 
of the form $(|v\rangle,{\bf 1},M)_B$ can perfectly fit the relative
frequencies of the outcomes (Proposition 5.1) and yet be mutually
orthogonal in the sense that $\langle v_\alpha| v_\beta \rangle = 0$
\end{prop}
\begin{prop} For measured data to uniquely decide to within unitary
equivalence which quantum-me\-chan\-ical model of a set of models best
fits experimental results interpreted as outcomes by a criterion of
least statistical distance (or any other plausible criterion), the set
of models must first be sufficiently narrowed, and this narrowing is
underivable from the results and the basic properties 1 and 2 of
quantum mechanics. \end{prop}

To these we now append:

\begin{prop} Any model $(|v \rangle, U, M)_B$ working in the
Hilbert space $\mathcal{H}$ can be mimicked exactly for all commands of
$B$ by two other mutually perpendicular models working with any command
set $\tilde{B} \supseteq B$ in the Hilbert space $\mathcal{H} \otimes
\mathcal{H}$. \end{prop}  
{\em Proof:} To generate one of the mimicking models, replace
$|v\rangle$ by $|\tilde{v} \rangle \otimes |w \rangle$ while for the
other use $|v\rangle \rightarrow |\tilde{v} \rangle \otimes |w_\perp
\rangle$, for any $|\tilde{v} \rangle|_B = |v\rangle$ and any $|w
\rangle, |w_\perp \rangle:\tilde{B} \rightarrow \mathcal{H}$ such that
$\langle w | w_\perp \rangle = 0 $; for both mimicking models, replace
$U$ by $U \otimes {\bf 1}$ and $M$ by $M \otimes {\bf 1}$.  An easy
check confirms that these models have the claimed properties of
producing the same probabilities for commands of $B$ as does the given
model, and that the two mimicking models are orthogonal to each
other. $\Box$

All the subsets of $S$ constitute a lattice under set intersection and
union; we call this a lattice of models.  Any record of measured
results interpreted as outcomes, together with a weighted statistical
distance $\epsilon$, defines a subset of $S$ consisting of those
models that have the property being within $\epsilon$ in distance to
the relative frequencies of the outcomes.  The propositions show that
$S$ is ``too big'' a set, in that it always contains models that are
drastically different from each other, indeed orthogonal.  That means
the subset contains models that, by any plausible measure, are
unappealing.

Each property applicable to $S$ defines a subset of $S$; {\em e.g.}
property 3 defines a subset, and property 4 a subset of that subset.
Indeed, the lattice of models can be viewed as isomorphic to a lattice
of properties.  Our scientist hopes to guess properties to produce a
subset within which the models that are close to measured results are
close to each other in the predictions they make, and have properties
that the scientist likes.  Actually, the scientist wants to guess
properties that narrow $S$ down to one model that appeals to the
scientist and best fits the measured results.

This can require a long journey.  Starting with $S$, the scientist
guesses a list of properties, each of which defines a subset.  The
list of properties defines a subset $S' \subset S$ that shares all
these properties, which is just the intersection of all the subsets
defined by the properties separately.  Having defined this smaller set
$S'$, the scientist explores the weighted distance between the models
of this subset and recorded measured data, and also explores the
closeness of the models that fit the data, if any, to each other.  If
carried out, this exploration results in one of three cases:
\begin{enumerate}
\item there are widely disparate models in $S'$ that all fit the data;
{\em i.e.} $S'$ is too big, so the scientist needs to guess additional
properties to shrink to fewer models;

\item no models of $S'$ fit well, leading to the need to drop
a property from the list to get a new subset of $S$, which
includes $S'$;

\item some models of $S'$ fit the data well and these are
close to each other in their implications, leading to the use of one
of these models, say model $\alpha$.
\end{enumerate}

\subsection{Ratcheting up precision takes fresh guesswork} 

Suppose the following: \begin{enumerate} \item A scientist cycles
through these cases, arriving eventually at case 3 with a model
$\alpha$, with $\epsilon$ as the weighted statistical distance between
the model and relative frequencies of outcomes obtained from the
testing of instruments. \item Using the model to define commands for
the CPC to transmit to instruments results in successful quantum
searching for functions of $n-1$ bits, but the commands do not work for
$n$ bits; per Eq. (11), requires a reduction of error by
$2^{-1/2}$. \item The scientist needs to produce commands that more
precisely generate behavior of the instruments described by quantum
gates specified as unitary matrices $U_j$ for $j = 1, \ldots , G$, and
needs to achieve weighted statistical distance $\epsilon' = 2^{-1/2}
\epsilon$, or less. \end{enumerate} 

\subsubsection{Can searching supplant the need for a better model?}
If a scientist has a command $b_j$ that generates behavior of
instruments consistent to within $\epsilon$ of a desired unitary
transformation $U_j$, can the scientist likely find a command that is
precise to within $\epsilon' = 2^{-1/2}\epsilon$ by searching, thus
skipping the need to get a more precise model?  That depends on the
how many commands have to tried, and the number of repetitions of
each command necessary to obtain a sample size big enough to
discriminate between models to within a statistical distance
of $\epsilon'$.

To determine this number of commands that have to be tried, we
introduce a notion of an $\epsilon$-grid on a metric space, here the
Lie group $SU(N)$ with the metric induced by the spectral norm.  By an
$\epsilon$-grid on a metric space $\mathcal{S}$, we mean a finite
subset $\mathcal{G}_\epsilon \subset \mathcal{S}$ such that for any $x
\in \mathcal{S}$, $\exists y \in \mathcal{G}_\epsilon$ such that $|x -
y| < \epsilon$.  To be sure of finding a command $\tilde{b}_j$ that
produces behavior within $\epsilon'$ of $U_j$, it is necessary to
explore all the points of an $\epsilon$'-grid of the $\epsilon$-ball
in $SU(N)$ around $U(b_j)$.  There are at least as many such points as
the ratio of the volume of an $\epsilon$ ball to the volume of an
$\epsilon$'-ball in $SU(N)$.  That ratio is $(\epsilon/ \epsilon')^d$,
where $d$ is the dimension of $SU(N)$, so $d = N^2 - 1 = 2^{2n} - 1$.
Thus the number of points that have to be explored to guarantee
finding one that improves the precision by enough to expand quantum
search by one bit is, for just one gate command, $2^{d/2}$, with $d
\approx 4^n$.  Starting with precision just adequate for a 4-bit
search, to reach precision just adequate for a 5-bit search requires
exploring about $10^{154}$ commands.  Without counting the large
number of trials needed to get a sample size big enough, blind
searching is hopeless as a way to find better commands.  Finding a
better model to generate commands, and thus avoiding a search over all
these commands, appears indispensable.

\subsubsection{Trying for a better model} A model $\alpha$ found in the
beginning of a quantum-computing experiment by trying out small
problems for a small sample sizes and a small set of commands $B_{\rm
test}$ will likely be inadequate to generate the precision required
to deal with larger problems, such as factorizing larger integers or
searching a larger list.  With blind searching ruled out, a better
model, call it $\beta$, is necessary from which to determine more
precisely the commands and their timing. 
 
How does looking for and deciding on a better model work?  Can this be
done by a universally applicable program, run by the CPC that controls
the quantum computing instruments?  Hope for this automation lies
mostly the idea that searching for a better model is adequately
defined by the objective of a better fit to measured results and by a
desire for simplicity in the model; the hope is perhaps spurred also
by the allure of the universal test used in mathematical logic,
isolated from laboratory instruments, for testing a claimed
derivation, a test that can be thought of as a universally applicable
program for a Turing machine\cite{boolos}.

This hope is dashed by Wittgenstein's contribution to the foundations
of both psychology and set theory, when he made unequivocally clear
that words are what people say in various scenes, so that what the
speaking of a word means depends on the scene in which the word is
spoken\cite{wittgenstein}.  This applies to words transmitted by a CPC
as commands to computer-controlled instruments.  The evaluation of the
words cannot be independent of the evaluation of the activity and
actions of the instruments generated by those words.  (Without this
connection any concept of testing falls in confusion, because the same
word, whether uttered by a person or transmitted as a command to an
instrument by a CPC, produces actions that depend on context beyond
the power of verbal or mathematical analysis, as follows from the
previously cited proof of the need for guesswork to link models to
instruments.)  This means that, even under our bold assumption that
suitable commands exist, the finding of a model with certain
properties that fits measured data at one level of precision by no
means assures the existence of a model having those properties and
fitting the same measured data at greater precision, let alone data
obtained from larger samples and a larger set of commands.  Thus when
larger problems are tackled, with their demand for increased
precision, the scientist can expect to have to go back through the
lattice of models, encountering all three cases, to find new, possibly
more complex properties by which to narrow the set of models.  This
implies a need for a scientist operating a quantum computer to switch
back and forth between modes of quantum computing and modes of
refining the model and the commands it calls for.\cite{ams}

\begin{figure}[h]
\epsfysize=4.375in\begin{center}
\hskip30pt\epsfbox{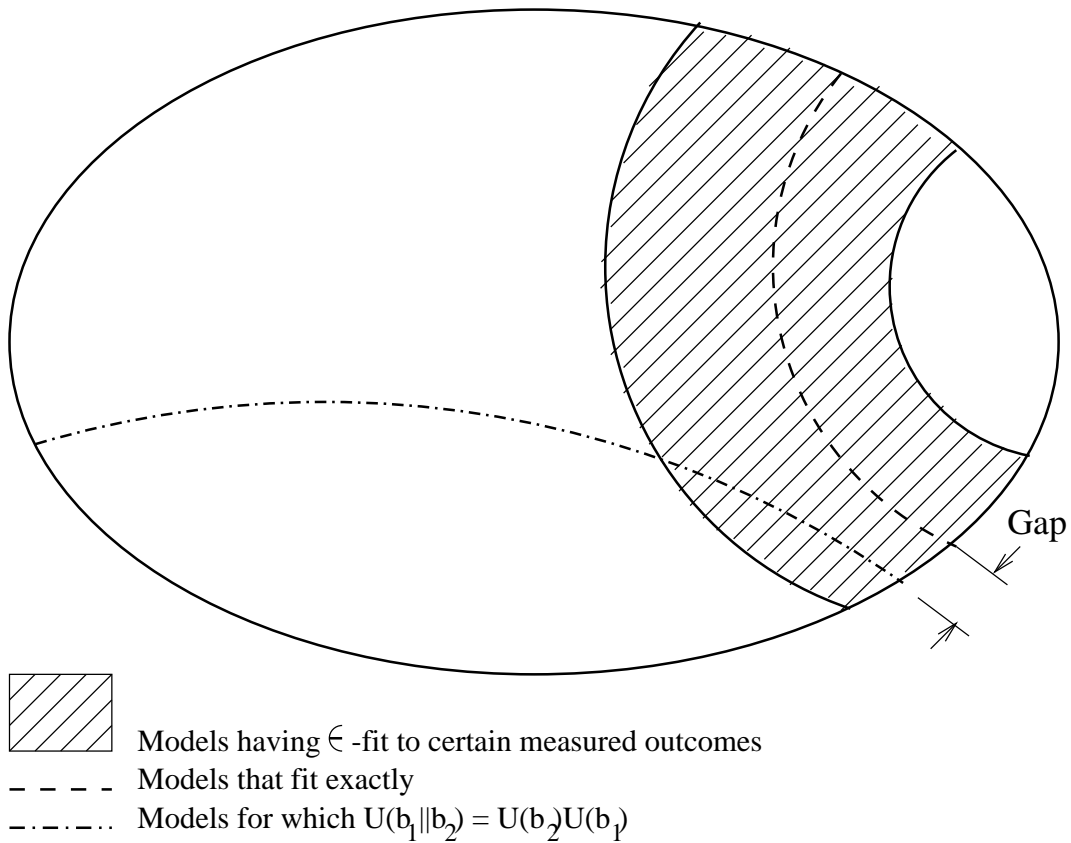}
\end{center}
\caption{Non-intersection of ``nice models'' with ``exactly fitting models.''}
\vskip-.5\baselineskip
\end{figure}

\section{Concluding remarks}

It has been striking and surprising to find in the course of this work
that contact between instruments and equations is a barely discovered
discipline of physics, awaiting further investigation.  The example of
quantum searching shows up some of the issues of contact especially
vividly.  The quantum computing problem of factorizing and the
Deutsch-Jozsa problem require a number of gate transformations
polynomial rather than exponential in the number of bits, and so the
challenges to their implementation are less extreme.  Nonetheless
implementing any device described quantum mechanically at progressive
levels of precision calls for navigating a lattice of models, coping
with ambiguity among inequivalent models, dealing with sample sizes
made necessary by quantum indeterminacy, and managing via machinery
that is specified classically the timing of signals at the precision
demanded by quantum mechanics.  Here we have provided some tools for a
promising discipline.  Since they are of general application, not
limited to quantum computing, these tools are widely applicable to
efforts that join theory and experiment in quantum physics.

\section*{Acknowledgments}
We thank Amr Fahmy for contributing to our thoughts and discussions at
many points during this work.  In addition to him, we are indebted to
Steffen Glaser, and Raimund Marx, and Wolfgang Bermel for introducing
us to the subtleties of laboratory work aimed at
nuclear-magnetic-resonance quantum computers\cite{marx}.  We learned
the view of quantum searching used here from David Mumford.

\end{document}